\documentclass[11pt]{article}
\usepackage{amsmath,amssymb}
\usepackage{epsfig,graphicx}
\usepackage{cite}

\def\id{\protect{{1 \kern-.28em {\rm l}}}}

\begin{document}

\title{\bf On nonlinear classical electrodynamics with an axionic term}
\author{E.~M.~Murchikova\footnote{{\bf e-mail}: e.murchikova@imperial.ac.uk}
\\
\small{\em Blackett Laboratory, Imperial College London, SW7
2AZ London, UK}\\
\small{\em Skobeltsyn Institute of Nuclear Physics, Moscow State University} \\
\small{\em 119991 Moscow, Russia} }

\date{}
\maketitle

\vspace{-7cm}
\begin{flushright}
\end{flushright}
 \vspace{-1cm}
\begin{flushright} Imperial-TP-EM-2010-01
\end{flushright}

\vspace{5cm}

\begin{abstract}

Recently there has been a renewed interest in axionic generalization
of electrodynamics due to its application to topological insulators.
A low-energy electromagnetic response of these exotic materials was
proposed to be described by an axionic term in the Lagrangian.
Motivated by this it is of interest to 
study various aspects of axionic electrodynamics and analyze the
universal  features of the axionic effects.
Here  we discuss the axionic modification of
 generalized electrodynamics  with a  Lagrangian being 
 an arbitrary function of two electromagnetic invariants.
Surprisingly,  the qualitative characteristics  of the major axionic effects
known in the Maxwell theory  happen to be  independent of  the exact type
of the nonlinear  Lagrangian  and are uniquely fixed by the form of the
axionic term.

\end{abstract}

\maketitle

Topological insulator (TI) is a recently discovered type of
materials. An anomalous band structure makes it insulating in the
bulk and an exotic metal on the surface and boundary (electrons on
its surfaces are insensitive to scattering by impurities). The first
reported example of such (2D) states is the quantum Hall effect,
where ``the metal'' occurs on the edges of the sample. Similar
behaviour was predicted to appear \cite{Fu:2007prb,Zhang:2009np} and
experimentally discovered \cite{Qi:2008prb,Xia:2009np,Chen:2009s} in
several 3D materials: $\rm Bi_{1-x} Sb_x$ alloys and $ \rm Bi_2
Se_3,$ $\rm Bi_2 Te_3,$ where the role of the magnetic field is
assumed by spin-orbital coupling and intrinsic properties of matter.
Exotic characteristics make TI a promising candidate for
applications in quantum computing, in particular for constructing
a quantum bit (for a review on TI see \cite{Moore:2010n}).

It was proposed, that a low-energy electromagnetic response of TI is
described by the axionic term in the Lagrangian \cite{Qi:2008prb}
\begin{equation}\label{Ltheta}
L_{\theta}=\theta ({\bf E} \cdot {\bf B}).
\end{equation}
Influence of this term on Maxwell electrodynamics was initially
studied many years ago
and several
fascinating properties of an interface between two media with different values
of
$\theta$ (an interface with a non-trivial jump
of $\theta$) were reported: magnetic charges induce electric mirror
charges and vice-versa \cite{Sikivie:1984pl}; a magnetic monopole
surrounded by a spherical axionic domain boundary picks up a
non-trivial electric charge \cite{Wilczek:1987prl}; reflection off
an axionic domain wall induces a non-trivial rotation of the
polarization of the fields \cite{Huang:1985pr,Obukhov:2005}. Some of
these phenomena are expected to be experimentally observable at the
surface of TI \cite{Qi:2009s}.

In light of the renewed interest in the axionic electrodynamics\footnote{
It  may be mentioned that the Lagrangian describing a real antiferromagnet
chromium oxide ${\rm Cr}_2{ \rm O}_3$ was also shown to have an axionic piece \cite{Hehl}.},
it is of interest to turn again to theoretical exploration
of various aspects of axionic electrodynamics
and analyze the
common features of the axionic affects not related to  particular
type of electrodynamics chosen. To this end we discuss axionic modification of
the generalized electrodynamics, the theory with the  Lagrangian
which is  an arbitrary function of two electromagnetic invariants.
Surprisingly,  the qualitative characteristics of the major axionic
effects expected on the basis of Maxwell theory
turn out to be  uniquely fixed by the form of the
axionic term and are  independent on the exact form  of the electrodynamics
Lagrangian. 

Before discussing  the generalized theory we consider 
a particular example of nonlinear electrodynamics
 --- the  Born--Infeld (BI) theory \cite{BornInfeld1934} (section 2).
We derive
the laws of refraction for electric and magnetic fields
at the interface with non-trivial jump
of $\theta$ and make use of them when considering electrically and
magnetically charged plane at the axionic boundary.
Section 3 is dedicated to a discussion of quantitative and
qualitative predictions of axionic effects in  generalized
electrodynamics with  Lagrangian as an arbitrary function
of two electromagnetic invariants.
In Appendix A 
we consider an
alternative approach to effective generalization of axionic electrodynamics
at the level of equations of motion.
 Depending on the way of treating magnetic monopoles (as genuine or
just an effective objects) there is an ambiguity
in  performing such a generalization. It results, for example, in opposite
predictions for  existence of the Witten effect in a specific configuration
of magnetic charges and axionic fields.

\section{BI electrodynamics with an axionic term}

\subsection{Maxwell equations for the BI axionic electrodynamics}

We start with  a detailed analysis of a particular example
of the nonlinear electrodynamics
 --- the  Born--Infeld (BI) theory \cite{BornInfeld1934}.

Similarly to the Maxwell theory, the BI axionic electrodynamic is
classically invariant under the $SL(2, {R})$ duality
transformations \cite{Gibbons:1995cv,Gibbons:1995ap,Tseytlin:1996np}
(see \cite{Karch:2009prl} for a discussion of this symmetry in the
context of TIs in Maxwell theory). Originally it was formulated
\cite{BornInfeld1934} as an extension of  the standard Maxwell
electromagnetism where self-energy of a point electric charge is
finite. Interest in this theory was revived as it was shown that the
BI-type Lagrangian appears in string theory (see
\cite{Tseytlin:1999dj} and references therein). In particular,  the
world volume action for a bosonic D3-brane is of BI type with the
axionic term accounting for the coupling to a Ramond--Ramond scalar field
background (see, e.g.,  \cite{Johnson:2003gi}).

The Lagrangian of BI electrodynamics with the axionic term and a
source reads
\begin{equation}\label{bi1}
\begin{array}{ll}
	\displaystyle L_{\rm BI + \theta}\\
	\displaystyle
	\quad =b^2\left( 1-\sqrt{1+\frac{F^{\mu\nu}
F_{\mu\nu}}{2 b^2}-\frac{(\,^\ast F^{\mu\nu} F_{\mu\nu})^2}{16 b^4}}
	\right)+ \frac{1}{4} \theta \,^\ast F^{\mu\nu} F_{\mu\nu}-A_\mu
	j_{\rm e}^\mu \\
	\displaystyle \quad =b^2\left( 1-\sqrt{1+ b^{-2} ({\bf{B}}^2-{\bf{E}}^2)-b^{-4}({\bf{E}\cdot\bf{B}})^2}
	\right)+\theta ({\bf{E}\cdot\bf{B}})-A_\mu j_{\rm e}^\mu.
\end{array}
\end{equation}
Here $F_{\mu\nu}=\partial_\mu A_\nu-\partial_\nu A_\mu,$ $\,^\ast
F^{\mu\nu}=-\frac{1}{2} e^{\mu\nu\rho\lambda}F_{\rho\lambda},$
$e^{\mu\nu\rho\lambda}$ is the totaly antisymmetric
tensor $(e^{0123}=1)$, $\frac{1}{4}F^{\mu\nu}
F_{\mu\nu}=\frac{1}{2}({\bf{B}}^2-{\bf{E}}^2),$ $\frac{1}{4}\,^\ast
F^{\mu\nu} F_{\mu\nu}=(\bf{E}\cdot\bf{B}),$
$j_{\rm e}^\mu=\{\rho_{\rm e},{\bf j}_{\rm e}\}$ is an electric four-current and $b$ plays the role of
electric field cut-off (if ${\bf B}=0,$ then $|{\bf E}|<b$)\footnote{
The value of $b$ is defined by
$b = \frac{E_{max}}{c},$
where
$E_{max}=\frac{e}{4 \pi \epsilon_0 r_0^2 }$
is the maximal field strength achieved in the Coulomb configuration,
$r_0 \approx r_e = \frac{e^2}{4 \pi \epsilon_0 m c^2}$ is the classical radius
of the electron. Explicitly $ b \approx 6.4 \times 10^{11} \frac{V \cdot \ sec}{m^2}.$
}.
 When
$\theta=\theta(x)$ is constant, the axionic term in (\ref{bi1}) is a
complete derivative:
\begin{equation}\label{ComplDeriv}
	L_{\theta} \sim \theta \partial_\mu \left( e^{\nu\nu\rho\lambda}
	A_\nu \partial_\rho A_\lambda \right)
\end{equation}
and does not effect equations of motion. Further we will discuss
varying $\theta(x).$

From (\ref{bi1}) and the Bianchi identity
\begin{equation}\label{Bianchi0}
	\partial_\mu \,^\ast F^{\mu\nu}=0,
\end{equation}
we obtain the set of Maxwell equations
\begin{equation}\label{bimaxgen}
	\left\{
	   \begin{array}{ll}
		 \displaystyle \nabla \cdot {\bf D}=\rho_{\rm e} \\
		 \displaystyle \nabla \cdot {\bf B}=0 \\
		 \displaystyle -\nabla \times {\bf E}=\frac{\partial {\bf B}}{\partial t}
		 \\[5pt]
		 \displaystyle \nabla \times {\bf H}=\frac{\partial {\bf D}}{\partial t}+{\bf j}_{\rm e} \\
	   \end{array}
	 \right.
\end{equation}
Here the electric displacement and magnetizing fields are defined
respectively as
\begin{equation}\label{dual2}
	\begin{array}{ll}
	  \displaystyle {\bf D} = \frac{\partial L_{\rm BI+\theta}}{\partial {\bf E}} \\
	  \displaystyle \quad \, = \frac{{\bf{E}}+{b^{-2}}({\bf{E}\cdot\bf{B}}) \, {\bf{B}}}
	  {\sqrt{1+ b^{-2} ({\bf{B}}^2- \,{\bf{E}}^2)-b^{-4}({\bf{E}\cdot\bf{B}})^2}}
	  +\theta {\bf B}={\bf D}_0+\theta {\bf B}, \\
	  \displaystyle {\bf H} = -\frac{\partial L_{\rm BI + \theta}}{\partial {\bf B}} \\
	  \displaystyle \quad \, = \frac{{\bf{B}}-{b^{-2}}({\bf{E}}\cdot{\bf{B}}) \, {\bf{E}}}
	  {\sqrt{1+ b^{-2} ({\bf{B}}^2- \,{\bf{E}}^2)-b^{-4}({\bf{E}\cdot\bf{B}})^2}}
	  -\theta {\bf E}={\bf H}_0-\theta {\bf E},
	\end{array}
\end{equation}
where the notations ${\bf D}_0$ and ${\bf H}_0$ stand for the fields
in the non-axionic BI theory.

\subsection{Refraction of electric and magnetic fields at the axionic domain boundary
}

Let us consider a planar axionic domain wall at $z=0:$
\begin{equation}\label{axwall2}
	\theta(z)=
	\begin{cases}
	\theta_1 \qquad {\rm for} \ z < 0\\
				\theta_2	\qquad {\rm for} \ z \ge 0
		\end{cases}
\end{equation}
Our aim is to find the laws of refraction for electric and magnetic fields
at such boundary --- an analogue of the Snell's laws for BI axionic
theory. Similarly to the optical Snell's laws which give
the value and the direction of
electric and magnetic fields on the other side of the interface with
different refractive indexes, ``axionic Snell's laws'' give
the value and the direction of
electric and magnetic fields on the other side of the interface with
different $\theta.$

In the absence of surface currents and charges Maxwell equations
(\ref{bimaxgen}) demand continuity for the perpendicular ($\bot$) to
the boundary components of ${\bf D}$ and ${\bf B}$ and the parallel
($\|$) components of ${\bf H}$ and ${\bf E}:$
\begin{equation}\label{biCont}
   \begin{array}{c}
	  \displaystyle {D}_\bot^{+}={D}_\bot^{-}, \qquad {B}_\bot^{+}={B}_\bot^{-}, \\
	  \displaystyle {H}_\|^{+}={H}_\|^{-}, \qquad
	  {E}_\|^{+}={E}_\|^{-},
	\end{array}
\end{equation}
where ``$-$'' and ``$+$'' refer to the fields behind ($z<0$) and in
front ($z>0$) of axionic wall, respectively. In the explicit form
(\ref{biCont}) read
\begin{equation}
	\begin{array}{ll}
	\displaystyle
	   B^+_\bot=B^-_\bot, \\
	\displaystyle
	   E^+_\|=E^-_\|, \\
	\displaystyle
	  \frac{{E^+_\bot}+{b^{-2}}({\bf{E^+}\cdot\bf{B^+}}) \, {B^+_\bot}}
	  {\sqrt{1+\, b^{-2} \, ({\bf{B^+}}^2-{\bf{E^+}}^2)- b^{-4}\, ({\bf{E^+}\cdot\bf{B^+}})^2}}
	  +\theta_2 B^+_\bot \\
	\displaystyle
	  \qquad \qquad \qquad=
	  \frac{{E^-_\bot}+{b^{-2}}({\bf{E^-}\cdot\bf{B^-}}) \, {B^-_\bot}}
	  {\sqrt{1+ \, b^{-2} \, ({\bf{B^-}}^2-{\bf{E^-}}^2) - b^{-4}\, ({\bf{E^-}\cdot\bf{B^-}})^2}}+\theta_1 B^-_\bot \, , \\
	\displaystyle
	  \frac{{B^+_\|}-{b^{-2}}({\bf{E^+}\cdot\bf{B^+}}) \, {E^+_\|}}
	  {\sqrt{1+ \, b^{-2} \, ({\bf{B^+}}^2-{\bf{E^+}}^2)- b^{-4}\, ({\bf{E^+}\cdot\bf{B^+}})^2}}
	  -\theta_2 E^+_\| \\
	\displaystyle
	  \qquad \qquad \qquad=
	  \frac{{B^-_\|}-{b^{-2}}({\bf{E^-}\cdot\bf{B^-}}) \, {E^-_\|}}
	  {\sqrt{1+ \, b^{-2} \, ({\bf{B^-}}^2-{\bf{E^-}}^2) - b^{-4}\, ({\bf{E^-}\cdot\bf{B^-}})^2}}-\theta_1
	  E^-_\|.
	\end{array}
\end{equation}

One may easily find the general solution of these equations. In
particular cases it looks as follows:

 if ${\bf B}^- \neq 0,\quad$ ${\bf E}^- = 0:$
\begin{equation}\label{bi5a-1}
	\begin{array}{ll}
		\displaystyle {\bf B}^+={\bf B}^-, \qquad E^{+}_\|=0, \\
		\displaystyle E^{+}_\bot=-\frac{\theta B^{-}_\bot}
		{\sqrt{1+b^{-2}{B^-_\bot}^2 (1+\theta^2)}}
		\sqrt{ \frac{1+b^{-2} {{\bf B}^-}^2}
		{1+b^{-2}{B^-_\bot}}	 };
	\end{array}
\end{equation}

 if ${\bf B}^- =0,\quad$ ${\bf E}^- \neq 0:$
\begin{equation}\label{bi5a-2}
	\begin{array}{ll}
		\displaystyle B^{+}_\bot=0, \qquad {\bf E}^+={\bf E}^- , \\
		\displaystyle B^{+}_\|=\frac{\theta	 E^{-}_\|}
		{\sqrt{ 1-b^{-2}{E^-_\|}^2 (1+\theta^2) }}
		\sqrt{ \frac{1-b^{-2}{{\bf E^-}}^2}
		{1-b^{-2}{E^-_\|}^2}};
	\end{array}
\end{equation}

if $B^-_\bot, E^-_\bot \neq 0,\quad$ $B^-_\|, E^-_\| = 0:$
\begin{equation}\label{bi5a-3}
	\begin{array}{ll}
		\displaystyle B^{+}_\bot=B^{-}_\bot, \qquad B^+_\|=0, \qquad E^+_\|=E^-_\|, \\
		\displaystyle E^{+}_\bot=\frac{	  \frac{E^-_\bot}{\sqrt{1-b^{-2}{E^-_\bot}^2}} - \frac{\theta B^-_\bot}{\sqrt{1+b^{-2}{B^-_\bot}^2}} }
		{\sqrt{ 1 + b^{-2} \left( { \frac{E^-_\bot}{\sqrt{1-b^{-2}{E^-_\bot}^2}} - \frac{\theta B^-_\bot}{\sqrt{1+b^{-2}{B^-_\bot}^2}} } \right)^2
		}};
	\end{array}
\end{equation}

if $B^-_\bot, E^-_\bot =0,\quad$ $B^-_\|, E^-_\| \neq 0:$
\begin{equation}\label{bi5a-4}
	\begin{array}{ll}
		\displaystyle B^+_\bot=0, \qquad E^{+}_\bot=0, \qquad E^+_\|=E^-_\|, \\
		\displaystyle B^{+}_\|=\frac{ \frac{B^-_\|}{\sqrt{1+b^{-2}{B^-_\|}^2}} + \frac{\theta E^-_\|}{\sqrt{1-b^{-2}{E^-_\|}^2}}  }
		{\sqrt{ 1 - b^{-2} \left( { \frac{B^-_\|}{\sqrt{1+b^{-2}{B^-_\|}^2}} + \frac{\theta E^-_\|}{\sqrt{1-b^{-2}{E^-_\|}^2}} } \right)^2
		}},
	\end{array}
\end{equation}
where $\theta=\theta_2-\theta_1.$

As an example of application of the BI axionic Snell's laws,
consider a uniformly electrically $\rho_{\rm e}$ and magnetically $\rho_{\rm m}$ charged plane
parallel to the wall (\ref{axwall2}) and located at $z=d<0.$ The
electric and magnetic fields at $z<0$ side of the wall are
\begin{equation}\label{EMwall}
	\begin{array}{ll}
	\displaystyle
	  B^-_\bot={\rho_{\rm m}}/{2}\,, \\
	\displaystyle
	  E^-_\bot=
	  \frac{\rho_{\rm e}}{\sqrt{4+b^{-2} ( \rho_{\rm m}^2+\rho_{\rm e}^2) }}\,.
	\end{array}
\end{equation}
Using (\ref{bi5a-3}) one finds the fields on the other $(z>0)$ side:
\begin{equation}\label{EMwall+}
	\begin{array}{ll}
	\displaystyle
	  B^+_\bot={\rho_{\rm m}}/{2}\,, \\
	\displaystyle
	  E^+_\bot=\frac{{\rho_{\rm e}-\theta\rho_{\rm m}}}
	  {\sqrt{4+b^{-2} \left( {\rho_{\rm m}}^2+ {(\rho_{\rm e}-\theta\rho_{\rm m})}^2 \right)
	  }}\,.
	\end{array}
\end{equation}
We see that due to the axionic wall ``shielding'' the plane
effectively picks up an additional electric charge density
$\rho_{{\rm e}}^{\rm add}=-\theta \rho_{\rm m},$ similarly to what
is predicted in the framework of the Maxwell theory.

Axionic Snell's laws (\ref{bi5a-1})-(\ref{bi5a-4}) are quite
useful in calculation of various phenomena at an axionic boundary,
however in most cases this approach is unnecessarily complicated. In
the next section we will use an easier and more general approach.

\section{Generalized non-linear electrodynamics with the axionic term}

\subsection{Maxwell equations for the generalized electrodynamics}

Let us consider a generalized electrodynamics \cite{Plebanski} 
with the axionic term and a source
\begin{equation}\label{Ar1}
	L_{\rm gen+\theta}=L_{\rm gen}+\frac{1}{4} \theta \,^\ast F^{\mu\nu} F_{\mu\nu}-A_\mu
	j^\mu_{\rm e},
\end{equation}
where $L_{\rm gen}$ is an arbitrary electrodynamics Lagrangian, as a
function of two electromagnetic invariants\footnote{Note, the
function $L_{gen}$ is not completely random, as it is assumed to be
the Lagrangian of a consistent electrodynamics.}.

Maxwell equations obtained from (\ref{Ar1}) have the same form as
(\ref{bimaxgen}) with the constitutive relations
\begin{equation}\label{genDH}
	\begin{array}{ll}
	  \displaystyle {\bf D} = \frac{\partial L_{\rm gen+\theta}}{\partial {\bf E}}
	  = \frac{\partial L_{\rm gen}}{\partial {\bf E}}+\theta {\bf B}={\bf D}_0+\theta {\bf B}, \\[12 pt]
	  \displaystyle {\bf H} = - \frac{\partial L_{\rm gen + \theta}}{\partial {\bf B}}
	  = -\frac{\partial L_{\rm gen}}{\partial {\bf B}}-\theta {\bf E}={\bf H}_0-\theta {\bf E}.
	\end{array}
\end{equation}
Here ${\bf D}_0$ and ${\bf H}_0$ are the electric displacement and
magnetizing fields in the $L_{\rm gen}$ non-axionic theory. Using these
notation one may rewrite (\ref{bimaxgen}) as follows
\begin{equation}\label{GenMaxWil}
\left\{
	\begin{array}{l}
	  \displaystyle	 \nabla \cdot {\bf{D}}_0=
	  \rho_{\rm e}-\nabla\theta \cdot {\bf{B}}	\\
	  \displaystyle \nabla \cdot {\bf{B}}=0 \\
	  \displaystyle \nabla \times {\bf{E}}=-\frac{\partial {\bf{B}}}{\partial t}
	  \\[7pt]
	  \displaystyle \nabla \times {\bf{H}}_0
	  =\frac{\partial {\bf{D}}_0}{\partial t} + {\bf{j}}_{\rm e}
	  +	 {\bf{B}} \frac{\partial}{\partial t} \theta
	  + \nabla\theta \times \bf{E}
	\end{array}
\right.
\end{equation}

Indeed, the primary objects in electrodynamics are ${\bf D}$ and
${\bf H}$ (not ${\bf D}_0,$ ${\bf H}_0$), so e.g. an electric charge
must be treated as a source of ${\bf D}$ field with $\theta {\bf B}$
as an inseparable part. However, so far as we treat magnetic charges
as an effective objects and $\nabla \cdot
{\bf{B}}=0$ is still applicable, only an electric charge is a source of an electric
field and one may reinterpret axionic terms simply as additional
charge and current densities\footnote{Note, that at an axionic
boundary constrains (\ref{biCont}) hold for ${\bf D}, \, {\bf H},$
while ${\bf D}_0$ and ${\bf H}_0$ experience jumps.} 
(see Appendix A
for a discussion of an alternative approach to effective generalization
of the Maxwell equations,
when magnetic charges are treated as genuine physical objects and
the use of $\nabla \cdot {\bf{B}}=0$ is not allowed).

\subsection{Uniformly charged plane}

Consider a uniformly $\rho_{\rm e}$ and $\rho_{\rm m}$ charged plane located at
$z=d<0$ at the axionic wall (\ref{axwall2}):
\begin{equation}\label{GenMaxWil2}
\left\{
	\begin{array}{l}
	  \displaystyle	 \nabla \cdot {{D}}_0=
	  \rho_{\rm e}\delta(z+d)-\theta \delta(0) {{B}}(0)	 \\
	  \displaystyle \nabla \cdot {{B}}=\rho_{\rm m}\delta(z+d) \\
	\end{array}
\right.
\end{equation}
From this equations one finds that
\begin{equation}\label{Wall23}
\left\{
	\begin{array}{l}
	B^+=\rho_{\rm m}/2, \\
	D^+_0=D^-_0-\theta \rho_{\rm m}
	\end{array}
\right.\end{equation}
These fields may be interpreted (see (\ref{GenMaxWil})) as the
charged plane effectively receives an additional electric charge
$-\theta \rho_{\rm m}/2.$ It is exactly the same result as we obtained in
the previous section in the particular case of BI electrodynamics
using the axionic Snell's laws. Here we generalized it to an
arbitrary linear by source axionic electrodynamics.

\subsection{Spherical axionic boundary}

Consider a magnetic charge $q_{\rm m}$ surrounded by spherical axionic
domain boundary of radius $r_0$ \cite{Wilczek:1987prl}:
\begin{equation}\label{WilTheta3}
	\theta{(r)}=
	\begin{cases}0 \qquad {\rm for} \ r < r_0\\
				  \theta \qquad {\rm for} \ r \ge r_0
	\end{cases}
\end{equation}
Integration of equations (\ref{GenMaxWil}) over $0 \leq r \leq R,$ $R>r_0,$ using
spherical symmetry of the problem, gives
\begin{equation}\label{Sph}
	D_0=-\frac{\theta q_{\rm m}}{R^2}.
\end{equation}
We see that a magnetic charge in this configuration of $\theta(r),$
effectively picks up an electric charge $q_{\rm e}=-\theta q_{\rm m}$ and become
a dyon. This result is true in an arbitrary electrodynamics.

\subsection{Time-dependent $\theta$}

Consider linearly time-dependent $\theta$ \cite{Rosenberg:2010ia}:
\begin{equation}\label{time}
	\theta(x)=\theta_0 t
\end{equation}
and a magnetic charge $\rho_{\rm m}=q_{\rm m} \delta({\bf r}):$
\begin{equation}\label{GenMaxWil3}
\left\{
	\begin{array}{l}
	  \displaystyle	 \nabla \cdot {\bf{D}}_0=
	  \rho_{\rm e}	\\
	  \displaystyle \nabla \cdot {\bf{B}}=q_{\rm m} \delta({\bf r}) \\
	  \displaystyle \nabla \times {\bf{E}}=-\frac{\partial {\bf{B}}}{\partial t}
	  \\[7pt]
	  \displaystyle \nabla \times {\bf{H}}_0
	  =\frac{\partial {\bf{D}}_0}{\partial t} + {\bf{B}} \frac{\partial}{\partial t}
	  \theta (x)
	\end{array}
\right.
\end{equation}
Taking the divergence of the fourth equation in (\ref{GenMaxWil3}),
one finds
\begin{equation}\label{7}
	 \frac{\partial}{\partial t} \, \rho_{\rm e} = - \theta_0 \nabla \cdot {\bf B}=
	- \theta_0 \, q_{\rm m} \delta ({\bf{r}}).
\end{equation}
Thus a magnetic charge in linearly increasing/decreasing with time
axionic background (\ref{time}) become a dyon linearly
increasing/decreasing its electric charge.

\subsection{Comments on mirror charges}

All axionic effects are strongly restricted by two constrains
following directly from Maxwell equations (\ref{GenMaxWil}).
Firstly, all axionic effects are related to $B_\bot$ and $E_\|.$
Secondly, the continuity relations require $B^-_\bot=B^+_\bot$ and
$E^-_\|=E^+_\|$ at the axionic boundary, which is free of charges
and currents. In these bounds axionic Snell's laws of any axionic
electrodynamics would be qualitatively the same. Thus the mirror
charge effects must be also very similar to ones in the Maxwell
case, i.e. electric charges would induce magnetic mirror charges and
vice-versa.

However, one may not demand the mirror charges to be point-like. In
particular, in BI electrodynamics an electric charge is point-like
in terms of ${\bf D}_0,$ but a charge density in terms of ${\bf E}.$
In this context, magnetic charges are expected to be represented by
charge densities, as the axionic term establishes relation between
${\bf B}$ and ${\bf E}.$

\section*{Conclusion}

In this paper we have demonstrated that nonlinear modifications of the
ordinary Maxwell electrodynamics does not change
the qualitative, and in some cases also the quantitative,
predictions of the influence of the axionic term.

\section*{Acknowledgments}

I would like to thank Arkady Tseytlin for many helpful discussions
and careful reading of the draft of this manuscript and Yuri Obukhov
for comments and criticism. This work is supported by a grant of
Dynasty Foundation and in part by a grant of President of the
Russian Federation for Leading Scientific Schools (Grant No.
SS---4142.2010.2).

\appendix

\section{Axionic electrodynamics with magnetic monopoles. \\ Effective generalization at the level
of equations of motion}

In this section we start from equations of motion for an arbitrary
axionic electrodynamics (\ref{bimaxgen}) and, following Dirac,
generalize Bianchi identity (\ref{Bianchi0}) to include into
consideration a magnetic four-current $j_{\rm m}^\nu=\{\rho_{\rm m}, {\bf j}_{\rm m}
\}:$
\begin{equation}\label{BianchiGen}
	\partial_\mu \,^\ast F^{\mu\nu}=j_{\rm m}^\nu.
\end{equation}
In this case the generalized Maxwell equations read
\begin{equation}\label{MaxGen}
	\left\{
	   \begin{array}{ll}
		 \displaystyle \nabla \cdot ({\bf{D}}_0+\theta {\bf B})=\rho_{\rm e} \\
		 \displaystyle \nabla \cdot {\bf B}=\rho_{\rm m} \\
		 \displaystyle -\nabla \times {\bf E}=\frac{\partial {\bf B}}{\partial t}+{\bf j}_{\rm m} \\
		 \displaystyle \nabla \times ({\bf{H}}_0-\theta {\bf E})=\frac{\partial }{\partial t} ({\bf{D}}_0+\theta {\bf B})+{\bf j}_{\rm e} \\
	   \end{array}
	 \right.
\end{equation}
where the constitutive relations are given by (\ref{genDH}) with
${\bf D}_0$ and ${\bf H}_0$ being the electric displacement and
magnetizing fields in the generalized non-axionic theory.
Here we treat magnetic monopoles as genuine physical objects, 
thus the use of $\nabla \cdot {\bf B}=0$
is not allowed.

A shift of $\theta$ by a constant no longer leaves Maxwell equations
(\ref{MaxGen}) invariant. We may question how the electrodynamics
with the additional axionic term ($\theta-$electrodynamics) is
different from the non-axionic one.

Consider a pure electric charge density $\rho_{\rm e}$ in a space with
non-trivial $\theta:$
\begin{equation}\label{Echarge}
	\left\{
	\begin{array}{ll}
	  \nabla \cdot ({\bf D}_0+\theta {\bf B})=\rho_{\rm e} \\
	  \nabla \cdot {\bf B}=\rho_{\rm m}=0
	\end{array}
	\right.
\end{equation}
From these equations one finds that the electric charge density is a
source of the same field configuration as in the non-axionic theory:
\begin{equation}\label{Echarge2}
	\left\{
	\begin{array}{ll}
	  \nabla \cdot {\bf D}_0=\rho_{\rm e} \\
	  \nabla \cdot {\bf B}=0
	\end{array}
	\right.
\end{equation}
Now consider a pure magnetic charge density $\rho_{\rm m}:$
\begin{equation}\label{Mcharge}
	\left\{
	\begin{array}{ll}
	  \nabla \cdot ({\bf D}_0+\theta {\bf B})=\rho_{\rm e}=0 \\
	  \nabla \cdot {\bf B}=\rho_{\rm m}
	\end{array}
	\right.
\end{equation}
From these equations one finds
\begin{equation}\label{Mcharge2}
	\left\{
	\begin{array}{ll}
	  \nabla \cdot {\bf D}_0=-\theta \rho_{\rm m} \\
	  \nabla \cdot {\bf B}=\rho_{\rm m}
	\end{array}
	\right.
\end{equation}
The magnetic charge density generates a field configuration, which
is equivalent to the one generated by a dyon in the non-axionic
electrodynamics, in other words $\rho_{\rm m}$ effectively induces an
electric charge density $\rho_{\rm e}^{\rm eff}=-\theta \rho_{\rm m}.$

Though non-trivial ${\bf D}_0,$ resulting from $\rho_{\rm m},$ is just an
effective response of the axionic medium, a probe electric charge
will interact with $\rho_{\rm m}$ via electric force. The latter, in fact,
runs us into a problem: the object called ``a magnetic charge'' may
not statically interact with an electric charge. In this sense, for
an observer located in a space with non-trivial $\theta$ (a
$\theta-$observer), $\rho_{\rm m}$ is not a magnetic, but a dyonic charge
density.

Concerning electric and magnetic currents, one reaches similar
conclusions:
\begin{equation}\label{Ecurrent}
	\left\{
	\begin{array}{ll}
	  -\nabla \times {\bf E}=0 \\
	  \nabla \times ({\bf{H}}_0-\theta {\bf E})={\bf{j}}_{\rm e}
	\end{array}
	\right.
	\quad
	\Rightarrow
	\quad
	\left\{
	\begin{array}{ll}
	  -\nabla \times {\bf E}=0 \\
	  \nabla \times {\bf{H}}_0={\bf{j}}_{\rm e}
	\end{array}
	\right.
\end{equation}
and
\begin{equation}\label{Mcurrent}
	\left\{
	\begin{array}{ll}
	  -\nabla \times {\bf E}={\bf{j}}_{\rm m} \\
	  \nabla \times ({\bf{H}}_0-\theta {\bf E})=0
	\end{array}
	\right. \,
	\quad
	\Rightarrow
	\quad
	\left\{
	\begin{array}{ll}
	  -\nabla \times {\bf E}={\bf{j}}_{\rm m} \\
	  \nabla \times {\bf{H}}_0 = -\theta {\bf{j}}_{\rm m}
	\end{array}
	\right.
\end{equation}
While the electric current is a source of the same fields as in the
non-axionic theory, the magnetic current ${\bf j}_{\rm m}$ generates the
field configuration equivalent to one generated by a dyonic current
with ${\bf j}^{\,\rm eff}_{\rm e}=-\theta {\bf j}_{\rm m}.$

Let us rewrite Maxwell equations (\ref{MaxGen}) in terms of new
four-currents $\tilde{j}_{\rm e,m},$ which are the pure sources of
electric and magnetic fields in $\theta-$
electrodynamics\footnote{See \cite{Karch:2009prl} for the
considerations of the same transformations in explicitly $SL(2,R)$
duality covariant form in the case of Maxwell electrodynamics}:
\begin{equation}\label{ChargesRedef}
	\begin{array}{ll}
	  {\tilde{j}}^\mu_{\rm e} \equiv \{{j}^\mu_{\rm e}, \,{j}^\mu_{\rm m}=0 \}, \\
	  {\tilde{j}}^\mu_{\rm m} \equiv \{{j}^\mu_{\rm e}={\theta{j}}^\mu_{\rm m}, \
	  {j}^\mu_{\rm m} \}.
	\end{array}
\end{equation}
The resulting equations are the same as in the non-axionic
electrodynamics:
\begin{equation}\label{MaxGen2}
	\left\{
	   \begin{array}{ll}
		 \displaystyle \nabla \cdot {\bf D}_0=\tilde{\rho}_{\rm e} \\
		 \displaystyle \nabla \cdot {\bf B}=\tilde{\rho}_{\rm m} \\ [5pt]
		 \displaystyle -\nabla \times {\bf E}=\frac{\partial {\bf B}}{\partial t}+{\tilde{\bf{j}}}_{\rm m}
		 \\ [5pt]
		 \displaystyle \nabla \times {\bf H}_0=\frac{\partial {\bf D}_0}{\partial t}+{\tilde{\bf{j}}}_{\rm e} \\
	   \end{array}
	 \right.
\end{equation}
Non-axionic and $\theta-$ electrodynamics are dynamically
indistinguishable, but have different definitions of a magnetic
four-current. Globally it means that, when $\theta$ is varying in
space, the differentiation of ``magnetic'' or ``dyonic''
four-currents is also varying in space with respect to the value of
$\theta$ at the concrete position of an observer\footnote{One must
be extremely carefully defining magnetic four-current in each
particular problem and always refer to the location of the observer.
To avoid necessity of working with redefined magnetic four-current,
one may always shift $\theta(x)$ in a way to put an observer into
$\theta=0$ part of space, where the definition takes the
conventional form.}.

Discuss now a magnetic charge $Q_{\rm m}$ surrounded by a spherical vacuum
bubble of radius $r_0$ (\ref{WilTheta3}) \cite{Wilczek:1987prl}.
To start, one must clarify the meaning of the magnetic charge $Q_{\rm m}.$
It must be approved to be a magnetic charge by an observer located
at $r>r_0$ (a $\theta-$observer), who is actually doing this
experiment. The object defined as a monopole in $\theta-$system is a
dyon
\begin{equation}\label{ChargeInside}
	Q_{\rm m}=\{ q_{\rm e}=\theta q_{\rm m}, \, q_{\rm m} \},
\end{equation}
where $q_{\rm e}, \, q_{\rm m}$ are non-axionic electric and magnetic charges.
Inserting (\ref{ChargeInside}) into the bubble, one finds
\begin{equation}\label{NoWilIn}
\begin{array}{ll}
	\displaystyle D^{in}=\theta\frac{q_m}{r^2}\\[5pt]
	\displaystyle B^{in}=\frac{q_m}{r^2}
\end{array}
\,\, \Rightarrow \,\,
\begin{array}{ll}
	\displaystyle D^{out}=D^{out}_0+\theta B^{out}=\theta\frac{q_m}{r^2}\\[5pt]
	\displaystyle B^{out}=\frac{q_m}{r^2}
\end{array}
\,\, \Rightarrow \,\, D_0^{out}=0.
\end{equation}
Here ${\rm in}$ and ${\rm out}$ are refereed to the fields inside and outside
the spherical axionic boundary, respectively.

The magnetic charge does not pick up an electric charge. This
conclusion is opposite to one made in \cite{Wilczek:1987prl}. The
reason is the definition of a monopole. The object that were put
inside the bubble in \cite{Wilczek:1987prl} is $q_{\rm m},$ which is a
magnetic monopole in non-axionic electrodynamics, but a dyon
$\{-\theta Q_{\rm m}, Q_{\rm m} \}$ for a true $\theta-$observer. Thus the
observer sees exactly the same dyon as were put inside the bubble in
the beginning. The latter practically means that, in this approach,
there is no Witten effect in this configuration.

\end{document}